%% file: main.tex
\documentclass{aa}

\usepackage{graphicx}
\usepackage{booktabs}
\usepackage[para]{threeparttable}
\usepackage{threeparttablex}
\usepackage{longtable}
\usepackage{xtab}
\usepackage{xcolor}
\defcitealias{lim2017}{L17}
\defcitealias{roberts2021_LOFARclust}{R21b}
\usepackage{txfonts}
\begin{document}

   \title{LoTSS jellyfish galaxies: II. Ram pressure stripping in groups versus clusters}


   \author{I.D. Roberts
          \inst{1}
          \and
          R.J. van Weeren\inst{1}
          \and
          S.L. McGee\inst{2}
          \and
          A. Botteon\inst{1}
          \and
          A. Ignesti\inst{3}
          \and
          H.J.A Rottgering\inst{1}
          }

   \institute{Leiden Observatory, Leiden University, PO Box 9513, 2300 RA
Leiden, The Netherlands\\
              \email{iroberts@strw.leidenuniv.nl}
         \and
             University of Birmingham School of Physics and Astronomy, Edgbaston, Birmingham, UK\\
        \and
            INAF- Osservatorio astronomico di Padova, Vicolo Osservatorio 5, IT-35122 Padova, Italy\\
             }

   \date{Received September 15, 1996; accepted March 16, 1997}


\abstract{Ram pressure stripping is a frequently cited mechanism for quenching galaxy star formation in dense environments.  Numerous examples of ram pressure stripping in galaxy clusters are present in literature; however, substantially less work has been focused on ram pressure stripping in lower mass groups, the most common galaxy environment in the local Universe.  In this work we use the LOFAR Two-metre Sky Survey (LoTSS) to search for jellyfish galaxies (i.e.\ galaxies with ram pressure stripped tails extending beyond the optical disk) in $\sim$500 SDSS groups ($z<0.05$), making this the most comprehensive search for ram pressure stripping in groups to date.  We identify 60 jellyfish galaxies in groups with extended, asymmetric radio continuum tails, which are found across the entire range of group mass from $10^{12.5} < M_\mathrm{group} < 10^{14}\,h^{-1}\,\mathrm{M_\odot}$.  We compare the group jellyfish galaxies identified in this work with the LoTSS jellyfish galaxies in clusters presented in \citet{roberts2021_LOFARclust}, allowing us to compare the effects of ram pressure stripping across three decades in group/cluster mass.  We find that jellyfish galaxies are most commonly found in clusters, with the frequency decreasing towards the lowest mass groups.  Both the orientation of observed radio continuum tails, and the positions of group jellyfish galaxies in phase space, suggest that galaxies are stripped more slowly in groups relative to clusters.  Finally, we find that the star formation rates of jellyfish galaxies in groups are consistent with `normal' star-forming group galaxies, which is in contrast to cluster jellyfish galaxies that have clearly enhanced star formation rates.  On the whole, there is clear evidence for ongoing ram pressure stripping in galaxy groups (down to very low group masses), though the frequency of jellyfish galaxies and the strength of ram pressure stripping appears smaller in groups than clusters.  Differences in the efficiency of ram pressure stripping in groups versus clusters likely contributes to the positive trend between quenched fraction and host halo mass observed in the local Universe.\\}


   \maketitle
%

\section{Introduction} \label{sec:intro}

Most galaxies in the local Universe are found in galaxy groups \citep[e.g.][]{geller1983,eke2005,robotham2011}, where groups are typically defined as systems with three or more member galaxies and total masses $<\!10^{14}\,\mathrm{M_\odot}$ \citep[e.g.][]{mamon2007,connelly2012}.  Given the large number of galaxies in these systems, understanding the impact of the group environment on galaxy properties is critical for understanding the evolution of galaxies in the local Universe.
\par
Compared to the low density field, galaxy groups host a higher proportion of red, passive, gas-poor, early type galaxies but groups still host more star-forming, gas-rich, late type galaxies than massive galaxy clusters \citep[e.g.][]{wilman2005,blanton2009,mcgee2011,wetzel2012,brown2017}.  This makes groups an intermediate environment between clusters and the field where environment has started to affect the properties of member galaxies, but not to the extent where groups are dominated by galaxies on the red sequence.  In fact, groups likely play a significant role in the build up of the cluster red sequence through the process of ``pre-processing'' \citep[e.g.][]{fujita2004}.  Specifically, since structure growth is hierarchical, massive galaxy clusters are assembled through mergers with galaxy groups that deposit new galaxies into the cluster.  Roughly half of present day cluster galaxies may have joined their cluster as part of a lower mass group \citep[e.g.][]{mcgee2009,delucia2012,bahe2013}.  Furthermore, galaxy quenched fractions around clusters are enhanced relative to the field even at the virial radius and beyond, consistent with an environmental effect on star formation prior to cluster infall \citep[e.g.][]{vonderlinden2010,wetzel2012,haines2015,roberts2017,bianconi2018,olave2018,roberts2019}.  Though it is important to note that some galaxies beyond the virial radius will not be infalling for the first time, but instead backsplashing after already passing pericentre \citep[e.g.][]{mahajan2011,oman2013}.  Disentangling the contribution between infalling galaxies and backsplash galaxies is critical for constraining the effects of pre-processing in the cluster outskirts.
\par
One key question is whether the dominant quenching mechanisms differ in groups compared to clusters.  Recently, many works have argued that ram pressure stripping (RPS) plays an important role in quenching star formation in galaxy clusters \citep[e.g.][]{muzzin2014,brown2017,vanderburg2018,maier2019,roberts2019,ciocan2020}.  Ram pressure can quench galaxies either by directly stripping cold, star-forming gas from the disk \citep{vollmer2012,jachym2014,lee2017,lee2018,jachym2019,moretti2020}, or by stripping the more diffuse atomic gas \citep{kenney2004,chung2007,chung2009,kenney2015,yun2019} which will leave the galaxy quenched once it exhausts its remaining molecular gas reserves.  In some examples of RPS, referred to as `jellyfish galaxies', tails (or `tentacles') of stripped material are observed trailing the galaxy opposite to the direction of motion \citep[e.g.][]{poggianti2017,boselli2018}.  The strength of ram pressure scales with $\rho_\mathrm{ICM} v^2$, where $\rho_\mathrm{ICM}$ is the density of the intracluster medium (ICM) and $v$ is the relative velocity between galaxies and the ICM.  On average, both the density of the ICM and galaxy velocities are higher in clusters than groups, therefore the strength of ram pressure will be stronger in massive clusters than lower mass groups.  This begs the question of whether or not ram pressure is strong enough in groups to efficiently strip gas from galaxies.
\par
There are some examples of RPS in groups in the literature, one being the starburst galaxy NGC 2276 in the NGC 2300 galaxy group.  NGC 2276 has a gas tail likely from RPS, though it is also tidally interacting with NGC 2300. The stripped tail is apparent in the radio continuum ($\sim\!1.4\,\mathrm{GHz}$, \citealt{davis1997}) and at X-ray wavelengths \citep{rasmussen2006,wolter2015}.  NGC 2276 also shows a bow shock front opposite to the tail, with elevated radio continuum emission, $\mathrm{H\alpha}$ emission, and a large number of bright X-ray sources along the leading edge \citep{davis1997,wolter2015}.  \citet{rasmussen2006} and \citet{wolter2015} conclude that ram pressure (along with viscous effects) is responsible for both the disturbed morphology and high star formation rate in NGC 2276. Another example of a group galaxy with a long X-ray tail is NGC 6872 in the Pavo Group. \citet{machacek2005} suggest that this $90\,\mathrm{kpc}$ tail could be a result of ram pressure and/or viscous stripping in the group environment.  A few more studies have found `comet-like' \textsc{Hi} morphologies for galaxies in groups \citep{bureau2002,mcconnachie2007}, which are likely being driven by RPS. In particular, a recent MeerKAT study of the Fornax A group \citep{kleiner2021} present evidence for 9 galaxies in the midst of being pre-processed prior to accretion onto the Fornax cluster. Some of these galaxies display \textsc{Hi} deficiencies as well as \textsc{Hi} morphologies consistent with RPS \citep{kleiner2021}.   Finally, evidence for RPS stripping in groups has also been presented in the form of gas disks which are truncated relative to the stellar component, consistent with RPS removing gas from the outside-in \citep{sengupta2007,vulcani2018}.
\par
These previous works show that RPS occurs in at least some galaxy groups, though the small number of galaxies identified thus far make it difficult to contrast the prevalence and effectiveness of RPS in groups versus clusters.  Recently, \citet{roberts2021_CFIS} have performed a search for ram pressure candidates in SDSS groups and clusters with optical imaging from the Canada-France Imaging Survey \citep{ibata2017}.  \citet{roberts2021_CFIS} identify $\sim\!30$ ram pressure candidates galaxies in groups ($M_\mathrm{halo} < 10^{14}\,\mathrm{M_\odot}$), but there still remain uncertainties related to the accuracy of ram pressure identifications from optical imaging alone, given that the stellar disk may not always be strongly perturbed by ram pressure.  In \citet{roberts2021_LOFARclust} (hereafter \citetalias{roberts2021_LOFARclust}) we presented a sample of $\sim$100 jellyfish galaxies in nearby ($z<0.05$) galaxy clusters, identified from 144 MHz radio continuum tails in the LOFAR Two-metre Sky Survey (LoTSS, \citealt{shimwell2017,shimwell2019}).  At 144 MHz, LOFAR \citep{vanhaarlem2013} is sensitive to synchrotron emission from cosmic rays accelerated by supernovae.  For galaxies experiencing strong ram pressure, these cosmic rays can be stripped out of the galaxy and detected as RPS tails in the radio continuum \citep[e.g.][]{gavazzi1987,murphy2009,chen2020}, giving reliable identifications of jellyfish galaxies.  The largest assets of LoTSS are its high resolution ($\sim\!6''$) and high sensitivity ($\sim\!100\,\mathrm{\mu Jy/beam}$) observations over extremely wide fields, which upon survey completion will include the entire northern extragalactic sky.  Such a uniform, wide field survey is ideal for completing a comprehensive search for jellyfish galaxies in low redshift groups.  Especially given the fact that jellyfish galaxies may be rarer in groups than clusters, meaning a search likely needs to cover a large number of groups in order to build a significant sample.
\par
The purpose of this work is twofold: (a) to perform a comprehensive search for RPS in galaxy groups and determine how common RPS is in groups compared to clusters, and (b) to test whether the properties of jellyfish galaxies in groups differ systematically from the properties of jellyfish galaxies in clusters. With the 144 MHz radio continuum from LoTSS, we identify 60 jellyfish galaxies across a sample of 498 SDSS galaxy groups.  This is far and away the most comprehensive search for jellyfish galaxies in groups to date.  In Section~\ref{sec:data} we describe the datasets that we use as well as the methods for identifying jellyfish galaxies.  In Section~\ref{sec:jellyfish_freq} we consider how the frequency of jellyfish galaxies depends on halo mass, ranging from low-mass groups to massive clusters.  In Section~\ref{sec:orbital_hist} we constrain the orbital histories of group and cluster jellyfish galaxies, both using tail orientations and positions in projected phase space.  In Section~\ref{sec:sfr} we test whether the star formation enhancement observed for LoTSS jellyfish galaxies in clusters \citepalias{roberts2021_LOFARclust} is also present for jellyfish galaxies in groups.  Finally, in Sections \ref{sec:disc_conc} \& \ref{sec:summary} we give a brief discussion and summarize the main conclusions from this work.  Throughout, we assume a $\mathrm{\Lambda}$ cold dark matter cosmology with $\Omega_M=0.3$, $\Omega_\Lambda=0.7$, and $H_0=70\,\mathrm{km\,s^{-1}\,Mpc^{-1}}$.


\section{Data \& methods} \label{sec:data}

\subsection{Group and cluster samples} \label{sec:grp_sample}

\begin{figure}
    \centering
    \includegraphics[width=\columnwidth]{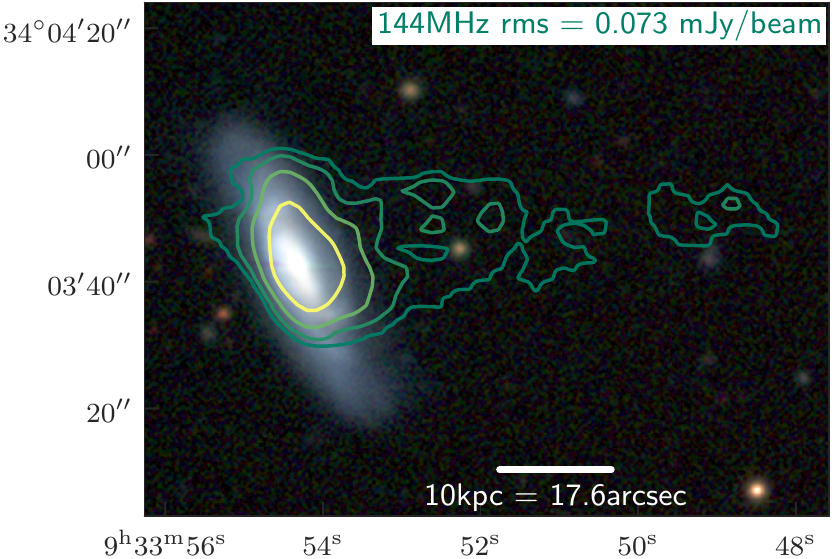}
    \caption{Optical $grz$ (DESI Legacy Survey, \citealt{dey2019}) image with LOFAR $144\,\mathrm{MHz}$ contours overlaid for KUG 0930+342, a jellyfish galaxy in a $1\times10^{13}\,\mathrm{M_\odot}$ galaxy group. Contours correspond to $2\times$, $4\times$, $8\times$, $16\times$, and $32\times$ the $144\,\mathrm{MHz}$ rms.}
    \label{fig:example_img}
\end{figure}

In this work we follow a similar methodology to \citetalias{roberts2021_LOFARclust} but focus on lower mass galaxy groups instead of galaxy clusters.  Our parent sample of galaxy groups comes from the \citet{lim2017} (hereafter \citetalias{lim2017}) SDSS group catalogue.  The \citetalias{lim2017} catalogue uses a group finder similar to that from the \citet{yang2005,yang2007} group catalogs but with improved halo mass estimates, especially for low mass systems.  Group masses in \citetalias{lim2017} are determined using abundance matching with a `halo mass proxy' that depends on both the stellar mass of the central galaxy and the stellar mass gap between the central galaxy and the $n$-th brightest satellite.  Comparisons to mocks show that this procedure typically reproduces the true halo masses without bias and with a typical uncertainty of $0.2\,\mathrm{dex}$ \citepalias{lim2017}.  From these halo masses, $M_\mathrm{halo}$, virial radii, $R_{180}$, and velocity dispersions, $\sigma$, for each group are estimated as \citepalias{lim2017}
\begin{equation}
    R_{180} = 1.33\,h^{-1}\,\mathrm{Mpc}\,\left(\frac{M_\mathrm{halo}}{10^{14}\,h^{-1}\,\mathrm{M_\odot}}\right)^{1/3} (1 + z_\mathrm{grp})^{-1}
\end{equation}
\noindent
and,
\begin{equation}
    \sigma = 418\,\mathrm{km\,s^{-1}}\,\left(\frac{M_\mathrm{halo}}{10^{14}\,h^{-1}\,\mathrm{M_\odot}}\right)^{0.3367}
\end{equation}
\noindent
For our group sample we include all groups from the \citetalias{lim2017} catalog that overlap with the $\sim\!5700\,\mathrm{deg^2}$ LoTSS DR2 (Shimwell et al. in prep.) footprint, and have: masses between $10^{12.5} < M_\mathrm{halo} < 10^{14}\,h^{-1}\,\mathrm{M_\odot}$, group redshifts of $z_\mathrm{grp}<0.05$, and galaxy memberships in the \citetalias{lim2017} catalogue of $N_\mathrm{galaxy}=5$ or more.  The redshift limit of $z<0.05$ is chosen to match that of the cluster sample in \citetalias{roberts2021_LOFARclust}, which allows us to make comparisons between the properties of jellyfish galaxies in groups versus clusters.

\begin{figure}
    \centering
    \includegraphics[width=\columnwidth]{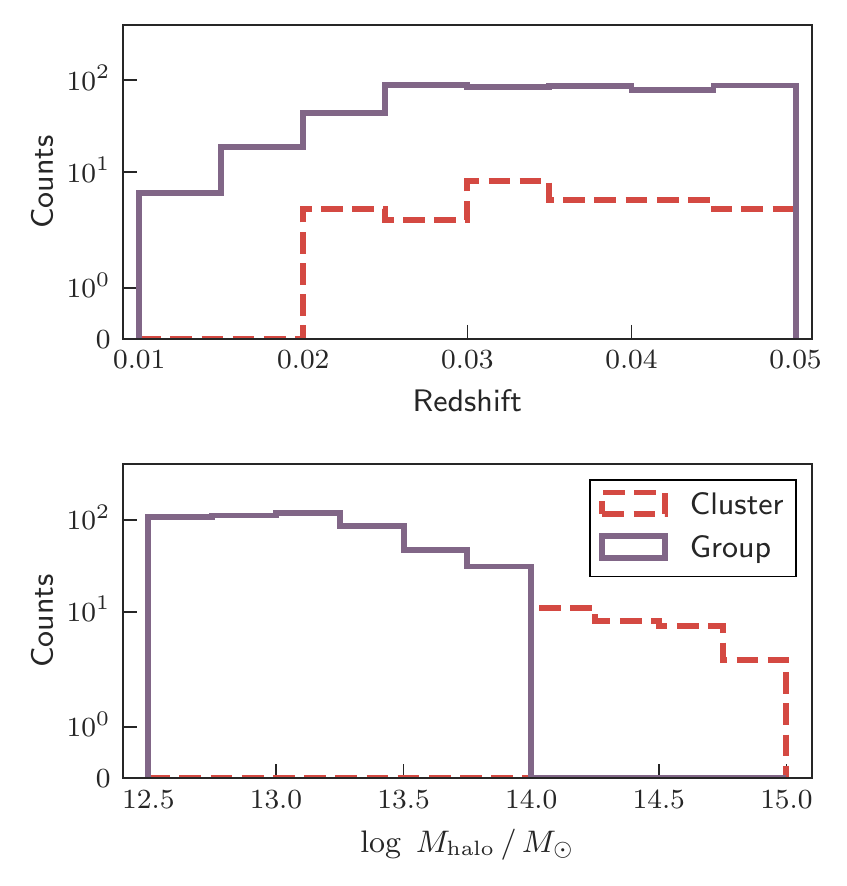}
    \caption{\textit{Top:} Redshift distribution for the sample of groups (purple, solid) and clusters (red, dashed).  \textit{Bottom:} Halo mass distribution for the sample of groups (purple, solid) and clusters (red, dashed).}
    \label{fig:z_Mh}
\end{figure}
\par
The \citetalias{roberts2021_LOFARclust} sample consists of 29 X-ray detected clusters from \citet{wang2014b} with $M_\mathrm{halo} \ge 10^{14}\,h^{-1}\,\mathrm{M_\odot}$, $z<0.05$, and have been observed by LOFAR at 144 MHz.  A detailed description of the cluster sample is given in \citetalias{roberts2021_LOFARclust}.  In Fig.~\ref{fig:z_Mh} we show the distribution of redshifts and halo masses for both the groups ($M_\mathrm{halo} < 10^{14}\,h^{-1}\,\mathrm{M_\odot}$) and the clusters ($M_\mathrm{halo} \ge 10^{14}\,h^{-1}\,\mathrm{M_\odot}$) in the sample.

\subsection{Galaxy samples} \label{sec:galaxy_sample}

\begin{table*}
    \centering
    \caption{Number of galaxies in various samples.}
    \begin{threeparttable}
    \begin{tabular}{l c c c c}
        \toprule
        Galaxy sample & Low-mass & Intermediate-mass & High-mass & Clusters\tnote{d} \\
        & groups\tnote{a} & groups\tnote{b} & groups\tnote{c} & \\
        \midrule
        SDSS galaxies & 1122 & 1371 & 1000 & 1968 \\
        LoTSS galaxies & 378 & 382 & 286 & 405 \\
        Jellyfish galaxies & 14 & 15 & 31 & 77 \\
        \bottomrule
    \end{tabular}
    \begin{tablenotes}
    \item[a] $10^{12.5} \ge M_\mathrm{halo} < 10^{13}\,\mathrm{M_\odot}$\\
    \item[b] $10^{13} \ge M_\mathrm{halo} < 10^{13.5}\,\mathrm{M_\odot}$\\
    \item[c] $10^{13.5} \le M_\mathrm{halo} < 10^{14}\,\mathrm{M_\odot}$\\
    \item[d] $M_\mathrm{halo} \ge 10^{14}\,\mathrm{M_\odot}$
    \end{tablenotes}
    \end{threeparttable}
    \label{tab:galaxy_samples}
\end{table*}

\subsubsection{Group member galaxies} \label{sec:group_galaxies}

For galaxies, we adopt a `loose' membership criteria (similar to \citealt{roberts2020}, \citetalias{roberts2021_LOFARclust}) where we include all galaxies that are within $1\times R_{180}$ of the stellar mass weighted group centre and $3\times \sigma$ of the group redshift as group members.  This ensures that we do not miss satellite galaxies at large velocity offsets, as is the case for many jellyfish galaxies \citep[e.g.][]{yoon2017,jaffe2018}.  Any galaxies that pass the membership criteria for multiple groups (this is only the case for <3\% of the galaxy sample) are assigned as members to the group that they are closest to in units of $R_{180}$.  To ensure a pure sample of galaxies in groups (i.e. $M_\mathrm{halo} < 10^{14}\,h^{-1}\,\mathrm{M_\odot}$), we also exclude any galaxies that are within $3 \times R_{180}$ in angular separation and $3000\,\mathrm{km\,s^{-1}}$ in redshift of any cluster in the \citetalias{lim2017} catalog (where we consider clusters to have $M_\mathrm{halo} \ge 10^{14}\,h^{-1}\,\mathrm{M_\odot}$).  For the galaxy sample we use stellar masses and star formation rates (SFRs) from the GSWLC-2 catalog \citep{salim2016,salim2018} that are determined by fitting galaxy SEDs with the \textsc{cigale} code \citep{boquien2019} that include UV, optical, and mid-IR fluxes.  This paper focuses on actively star-forming galaxies which we define as those galaxies with specific star formation rates $>\!10^{-11}\,\mathrm{yr^{-1}}$ (where, $\mathrm{sSFR} = \mathrm{SFR} / M_\mathrm{star}$).  In total, the above selections amount to a sample of 3493 star-forming `SDSS group galaxies' across 498 groups.
\par
From this sample of SDSS group galaxies, we use the forthcoming LoTSS DR2 source catalog (see \citealt{williams2019} for a description of the public LoTSS DR1 source catalogs) to find those galaxies that are also detected in LoTSS at 144 MHz.  We cross match the positions of SDSS group galaxies with the positions of LoTSS sources and keep any matches with separations $<\!3''$, which corresponds to the HWHM of the LoTSS beam.  This gives a sample of 1048 star-forming group galaxies with LoTSS detections, and we will refer to these galaxies as `LoTSS group galaxies'.

\subsubsection{Cluster member galaxies} \label{sec:cluster_galaxies}

The same membership criteria of $R < 1 \times R_{180}$ and $\Delta v < 3 \times \sigma$ is applied to the cluster sample from \citetalias{roberts2021_LOFARclust}, which gives 1968 star-forming `SDSS cluster galaxies' in 29 clusters ($M_\mathrm{halo} \ge 10^{14}\,\mathrm{M_\odot}$).  Star formation rates and stellar masses for cluster galaxies are also taken from the \citet{salim2016,salim2018} catalogue.  Star-forming SDSS cluster galaxies are cross matched with the LoTSS source catalog in the same way as for group galaxies.  This gives a sample of 405 `LoTSS cluster galaxies'.  In Table~\ref{tab:galaxy_samples} we summarize the size of the SDSS galaxy sample, the LoTSS galaxy sample, and the Jellyfish galaxy sample as a function of host halo mass.

\subsubsection{Field galaxies} \label{sec:field_galaxies}

We also construct a sample of isolated `field' galaxies.  The field sample consists of all galaxies in single-member groups from the \citetalias{lim2017} catalogue with $M_\mathrm{halo} < 10^{12.5}\,\mathrm{M_\odot}$ (i.e. consistent with an individual galaxy halo) and $z < 0.05$.  We then apply an isolation criteria (similar to \citealt{roberts2017}) and only include galaxies which are separated by at least $1000\,\mathrm{kpc}$ and $1000\,\mathrm{km\,s^{-1}}$ from the nearest galaxy with $M_\mathrm{star} \ge 10^{9.7}\,\mathrm{M_\odot}$.  $M_\mathrm{star} = 10^{9.7}\,\mathrm{M_\odot}$ corresponds to the SDSS stellar mass completeness at $z=0.05$ (\citealt{weigel2016}; \citetalias{roberts2021_LOFARclust}), therefore by only considering galaxy neighbours with $M_\mathrm{star} \ge 10^{9.7}\,\mathrm{M_\odot}$ we ensure that the strictness of this isolation criteria is independent of redshift (over the redshift range of our sample).  That said, it does mean that the galaxies in our field sample may not be isolated with respect to galaxies with stellar masses below this limit -- though we reiterate that none of the galaxies in the field sample were assigned to a group by the \citetalias{lim2017} algorithm.
\par
These criteria give a sample of 8044 star-forming SDSS field galaxies.  Again, matching these galaxies to sources in the LoTSS DR2 source catalog within $3''$ gives 2274 `LoTSS field galaxies'.

\subsection{Jellyfish galaxy selection} \label{sec:jellyfish_selection}

We take a two step approach to identifying jellyfish galaxies.  First, an automated pre-selection of `jellyfish candidates', and then second, by-eye classifications on all of the jellyfish candidates.  We pre-select jellyfish candidates with the shape asymmetry parameter ($A_S$, \citealt{pawlik2016}) applied to the LoTSS 144 MHz maps for all LoTSS group galaxies.  The shape asymmetry measures the rotational asymmetry of the binary detection maps (segmentation maps) for sources, and is calculated as
\begin{equation}
    A_S = \frac{\sum | X_0 - X_{180} |}{2\times\sum | X_0 |},
\end{equation}
\noindent
where $X_0$ is the source segmentation map and $X_{180}$ is the segmentation map rotated by $180^\circ$.  The shape asymmetry is a non-flux-weighted version of the commonly used CAS asymmetry \citep{abraham1996,conselice2003}, making it particularly sensitive to low surface brightness features such as ram pressure stripped tails.
\par
For each LoTSS group galaxy we create 144 MHz segmentation maps with the \texttt{photutils.detect\_sources} function in \textsc{Python} with a $3\sigma$ threshold.  We then pre-select jellyfish candidates as all LoTSS group galaxies with $A_S > 0.3$.  \citetalias{roberts2021_LOFARclust} show that this threshold of $A_S > 0.3$ includes $\sim\!85$\% of visually identified LoTSS jellyfish galaxies in clusters, while excluding $\sim\!70$\% of LoTSS sources in clusters which are not identified as jellyfish.  This pre-selection gives 271 jellyfish candidates which we then visually inspect to build our final sample of LoTSS jellyfish galaxies in groups.  We also include all LoTSS field galaxies which have $A_S > 0.3$.  `True' field galaxies should not be affected by RPS, so including field galaxies acts as a test of the methodology.  For the visual classifications we include field galaxies with $A_S>0.3$ randomly alongside group galaxies with $A_S>0.3$, such that the classifier does not know whether they are inspecting a group galaxy or a field galaxy.  Therefore if we are effective at selecting jellyfish galaxies associated with RPS in dense environments, very few field galaxies should pass this visual inspection.
\par
For visual inspections we follow \citetalias{roberts2021_LOFARclust} and make $100\,\mathrm{kpc} \times 100\,\mathrm{kpc}$ $g$-band cutout images from PanSTARRS and overlay 144 MHz flux contours from LoTSS.  LoTSS contours are only shown above $2 \times \mathrm{rms}$, where the rms noise is estimated locally from the LoTSS cutouts with sigma-clipped statistics. As in \citetalias{roberts2021_LOFARclust} we identify jellyfish galaxies as star-forming group galaxies which show `144 MHz emission which is resolved and clearly asymmetric with respect to the stellar disk of the galaxy (as traced by the g-band flux)'.  We reiterate that we only visually inspect galaxies with $A_S>0.3$, and we only inspect star-forming galaxies and therefore do not expect strong contamination from AGN emission \citepalias{roberts2021_LOFARclust}. We also note that our selection is not sensitive to galaxies with stripped tails along the line-of-sight, as such galaxies may not show clearly asymmetric radio continuum emission when projected in the plane of the sky. This is a source of incompleteness for our sample that is not easily remedied with imaging data alone.  Finally, any galaxies that show clear signatures of galaxy-galaxy interactions in their optical images are not included in the jellyfish sample, the same is true for for galaxies with close companions on the sky that are at the same redshift as the primary galaxy. This is done to limit the galaxies selected with tails due to tidal interactions as opposed to RPS. While we cannot say that our sample is completely free of such cases, the results of this work, and of \citetalias{roberts2021_LOFARclust}, are consistent with RPS being the primary driver of tail production in these galaxies.  Of the 271 jellyfish candidates in groups, 60 are identified as jellyfish galaxies through visual inspection.  This is the largest sample of RPS galaxies in groups identified to date.  In Fig.~\ref{fig:example_img} we show an example optical+radio image of a jellyfish galaxy in a $10^{13.1}\,\mathrm{M_\odot}$ group, where we have overlaid the LoTSS 144 MHz flux contours.  We show the PanSTARRs+LoTSS overlay images for all of the group jellyfish galaxies in Appendix~\ref{sec:img_appendix}.
\par
Of the LoTSS field galaxies, 2\% were classified as `jellyfish galaxies' by visual inspection.  While this is a non-zero fraction, the proportion of `jellyfish galaxies' in the field sample is clearly below that for the group and cluster samples (see Fig.~\ref{fig:jellyfish_Mh}).  Some of these field galaxies may be true jellyfish galaxies in small groups which have been mis-classified by the \citetalias{lim2017} group finder.  Alternatively, RPS may be possible, to some extent, in cosmic filaments \citep[e.g.][]{edwards2010,benitez-llambay2013} which could encompass some of our field sample.  Incorrect source association or emission from AGN could also give rise to asymmetric 144 MHz emission in field galaxies.  The purpose of this exercise is not to explain the origin of these `jellyfish galaxies' in the field sample (though there are plausible explanations, see above), but instead to get a sense of the false-positive rate of these visual inspections and understand the limits of this technique.
\par
Finally, we also include the sample of LoTSS cluster jellyfish galaxies from \citetalias{roberts2021_LOFARclust}.  These jellyfish galaxies are also identified from visual inspections in an analogous fashion to the group sample.  We only include jellyfish galaxies from \citetalias{roberts2021_LOFARclust} with $A_S>0.3$ (where $A_S$ is measured using the exact method described above) to ensure homogeneity with the group jellyfish galaxies in this work.


\section{How common are jellyfish galaxies in groups versus clusters?} \label{sec:jellyfish_freq}

\begin{figure}
    \centering
    \includegraphics[width=\columnwidth]{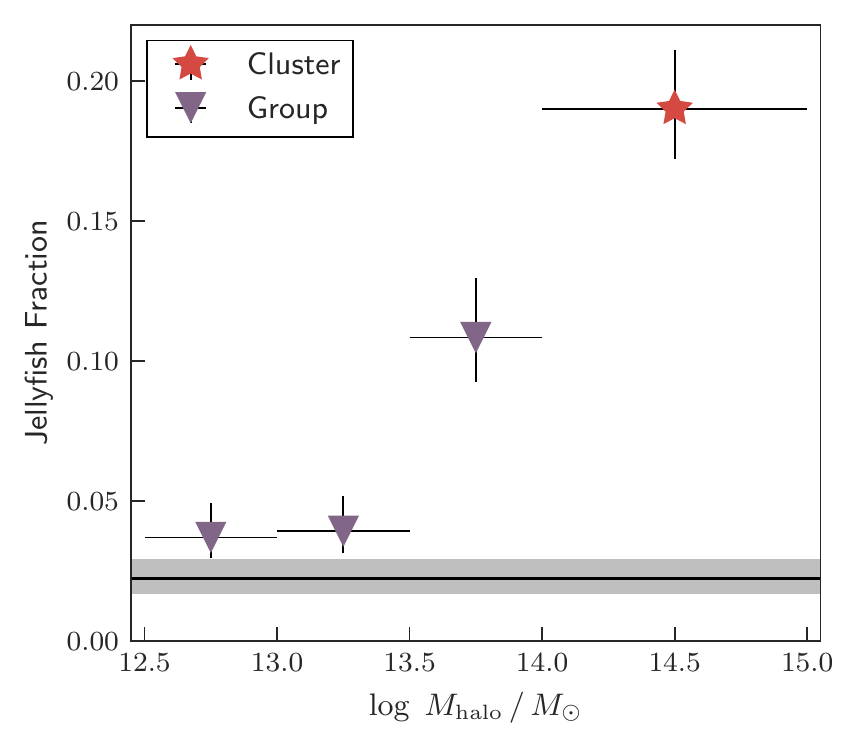}
    \caption{The jellyfish galaxy fraction (relative to all star-forming LoTSS sources) as a function of group/cluster halo mass.  Purple triangles show the jellyfish galaxies in groups identified in this work and the red star shows the cluster jellyfish galaxies from \citetalias{roberts2021_LOFARclust}.  Vertical error bars are 68\% binomial confidence intervals from \citet{cameron2011} and horizontal error bars show the width of each halo mass bin.  The horizontal line shows the fraction of LoTSS sources in the field sample that passed our jellyfish galaxy criteria (see Sect.~\ref{sec:jellyfish_selection}), along with the 90\% confidence region (shaded band).}
    \label{fig:jellyfish_Mh}
\end{figure}

In galaxy clusters, on average, both the ICM density and the relative velocities are larger than for groups (for simplicity, we use 'ICM' to refer to both the intra-cluster medium and the intra-group medium), therefore ram pressure stripping should be most prevalent in the cluster environment.  With the large sample of jellyfish galaxies that we have identified in groups, we can directly test this prediction.
\par
In Fig.~\ref{fig:jellyfish_Mh} we plot the fraction of jellyfish galaxies as a function of halo mass, for low-mass groups ($10^{12.5} \le M_\mathrm{halo} < 10^{13}\,h^{-1}\,\mathrm{M_\odot}$), intermediate-mass groups ($10^{13} \le M_\mathrm{halo} < 10^{13.5}\,h^{-1}\,\mathrm{M_\odot}$), high-mass groups ($10^{13.5} \le M_\mathrm{halo} < 10^{14}\,h^{-1}\,\mathrm{M_\odot}$), and galaxy clusters ($M_\mathrm{halo} \ge 10^{14}\,h^{-1}\,\mathrm{M_\odot}$, \citetalias{roberts2021_LOFARclust}).  The jellyfish galaxy fraction, $F_\mathrm{jellyfish}$, is defined for each halo mass bin as
\begin{equation} \label{eq:Fjelly}
    F_\mathrm{jellyfish} = \frac{N_\mathrm{jellyfish}}{N_\mathrm{LoTSS}}
\end{equation}
\noindent
where $N_\mathrm{jellyfish}$ is the number of LoTSS jellyfish galaxies and $N_\mathrm{LoTSS}$ is the number of star-forming galaxies detected in LoTSS.  We define the jellyfish fractions relative to the number of LoTSS sources in each halo mass bin instead of the number of SDSS member galaxies in each halo mass bin, due to the different stellar mass completeness between SDSS and LoTSS.  The majority of star-forming low-mass galaxies ($M_\mathrm{star} \lesssim 10^{9.5}\,\mathrm{M_\odot}$) in SDSS fall below the sensitivity limit of LoTSS (see \citetalias{roberts2021_LOFARclust} for a more complete discussion), therefore by defining the jellyfish fraction relative to LoTSS sources we are ensuring that both the numerator and denominator in Equation~\ref{eq:Fjelly} have similar stellar mass completeness.  That said, we have confirmed that when defining $F_\mathrm{jellyfish}$ in terms of SDSS group/cluster galaxies instead of LoTSS group/cluster galaxies, the qualitative trend shown in Fig.~\ref{fig:jellyfish_Mh} still holds.  Therefore our choice of denominator in Equation~\ref{eq:Fjelly} is not driving the results from this section.
\par
In Fig.~\ref{fig:jellyfish_Mh} we see that the jellyfish fraction steadily increases with halo mass, with a factor of $\sim\!4$ difference between low-mass groups and galaxy clusters.  This indeed suggests that ram pressure stripping is more prevalent in more massive halos.  The stellar mass distribution for LoTSS sources is very similar across the halo mass bins in Fig.~\ref{fig:jellyfish_Mh}, therefore it is unlikely that the observed trend with halo mass is being influenced by any stellar mass biases.  The trend levels off for low-mass groups, as the jellyfish fraction is similar in each of the two lowest halo mass bins.  We note that halo mass uncertainties will be highest for the lowest mass groups, this could lead to systems artificially scattering between the two lowest mass bins, which may contribute to the lack of observed trend for those masses. For all halo mass bins the jellyfish fraction is larger than the ``false-positive'' rate of 2\% that we find from the field sample, though the jellyfish fractions for the lowest-mass halos do come close this value. This suggests that while RPS does occur even in these very low mass groups, the vast majority of star-forming galaxies in such systems are not strongly affected.  The results in Fig.~\ref{fig:jellyfish_Mh} are consistent with previous works finding a higher fraction of galaxies undergoing RPS in more massive halos.  For ram pressure candidates identified from rest-frame optical imaging, \citet{roberts2021_CFIS} find a factor of two increase in the frequency of ram pressure candidates from groups to clusters, and in the Illustris simulation, \citet{yun2019} find a similar halo mass trend for simulated jellyfish galaxies.  While the methodologies for identifying RPS galaxies in these studies differ from this work, the qualitative trends between lower mass groups and massive galaxy clusters are consistent throughout.


\section{Orbital histories} \label{sec:orbital_hist}

Given the differences in velocity dispersions and ICM densities between low-mass groups and high-mass clusters, it is natural to expect that ram pressure stripped galaxies in groups may have different orbital histories than ram pressure stripped galaxies in clusters.  Previous work on jellyfish galaxies in clusters suggest that these objects begin to be stripped shortly after infall, before reaching the pericentre of their orbit (e.g. \citealt{yoon2017}; \citealt{jaffe2018}; \citetalias{roberts2021_LOFARclust}). Given weaker ram pressure in the group regime, there may be a substantial delay between galaxy infall and the onset of stripping, which is not seen in clusters.  In this section we constrain the orbital histories of jellyfish galaxies in both groups and clusters using two observational tools, the orientation of stripped tails with respect to the cluster centre and the position of galaxies in projected phase space.

\subsection{Tail orientations} \label{sec:tail_orient}

\begin{figure}
    \centering
    \includegraphics[width=\columnwidth]{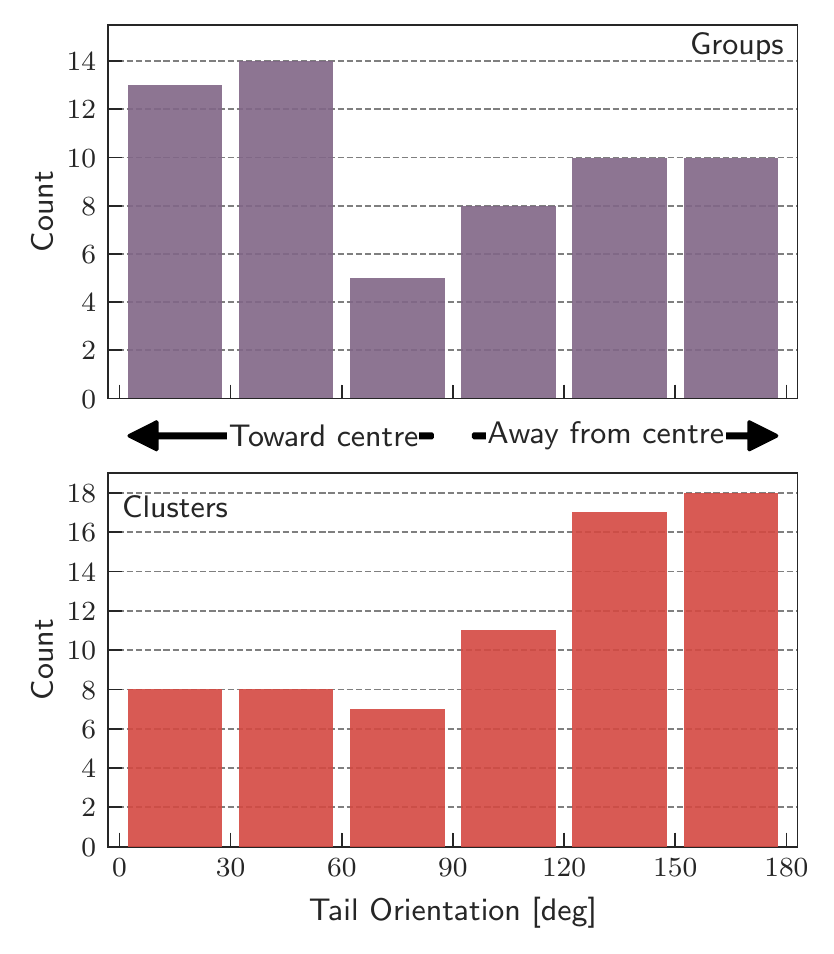}
    \caption{Orientation of jellyfish tails with respect to the cluster centre for groups (top) and clusters (bottom).  Orientations of $0^\circ$ correspond to tails aligned toward the cluster centre and orientations of $180^\circ$ correspond to tails aligned away from the cluster centre.}
    \label{fig:tail_orientation}
\end{figure}

In Fig.~\ref{fig:tail_orientation} we show the distributions of jellyfish tail orientations in groups (top) compared to clusters (bottom).  For both panels tail directions are measured with the same technique (see \citealt{roberts2020}; \citetalias{roberts2021_LOFARclust}), namely, for each $100\,\mathrm{kpc} \times 100\,\mathrm{kpc}$ PanSTARRs+LOFAR overlay image, the direction of the 144 MHz tail with respect to the optical galaxy centre is given an angle between $0^\circ$ and $360^\circ$.  The vector along this tail direction is then compared to the vector between the optical galaxy centre and the stellar mass weighted group centre, which gives a tail orientation relative to the group centre.  A tail pointing directly toward the group centre corresponds to an orientation of $0^\circ$ and a tail pointing directly away from the group centre corresponds to an orientation of $180^\circ$.
\par
In Fig.~\ref{fig:tail_orientation} differences are apparent between the distributions of tail orientations for jellyfish galaxies in groups (top) versus clusters (bottom).  For clusters, as shown in \citetalias{roberts2021_LOFARclust}, the distribution is clearly peaked at orientations between $120^\circ$ and $180^\circ$, consistent with galaxies being mostly stripped on first infall toward the cluster centre.  For groups, the distribution instead peaks most strongly at tail orientations $<\!60^\circ$.  This shows that many jellyfish galaxies in groups have tails oriented toward the cluster centre, consistent with galaxies on orbiting away from the centre after a pericentric passage.  There is also a significant number of group jellyfish with tail orientations between $120^\circ$ and $180^\circ$, suggestive of a mix of jellyfish galaxies on first infall and jellyfish galaxies backsplashing in the group environment.  This interpretation implies that jellyfish galaxies in groups have, on average, longer times-since-infall than jellyfish galaxies in clusters.  A natural explanation for this is the stronger ram pressure in clusters, capable of stripping galaxies relatively quickly after infall.  Whereas the onset of stripping in groups may be delayed due to lower ICM densities and galaxy velocities, for example, \citet{oman2021} estimate that groups strip satellites on timescales that are $\sim\!3\,\mathrm{Gyr}$ longer than for clusters based on observed star formation and \textsc{Hi} properties.  The orientations in Fig.~\ref{fig:tail_orientation} hint at this picture, but there are also complications related to the interpretation of such distributions, including projection effects and uncertainties around galaxy orbital parameters.  We also note that the tail orientations for the cluster sample are measured with respect to the X-ray centre, whereas tail orientations for the group sample are measured with respect to the stellar mass weighted group centre which is likely a less reliable tracer of the true minimum of the potential well.  X-ray centres are only available for a small fraction of our group sample therefore using a centre estimate based on galaxy positions is the only way, despite the added uncertainties.  Below we consider the distributions of group and cluster jellyfish galaxies in projected phase space (PPS), which is another tool to gain insight into group/cluster infall histories.  Specifically, we test whether the phase space distributions are consistent with the picture suggested by the tail orientations; namely, longer times-since-infall for jellyfish galaxies in groups versus clusters.

\subsection{Projected phase space} \label{sec:phase_space}

\begin{figure*}
    \centering
    \includegraphics[width=\textwidth]{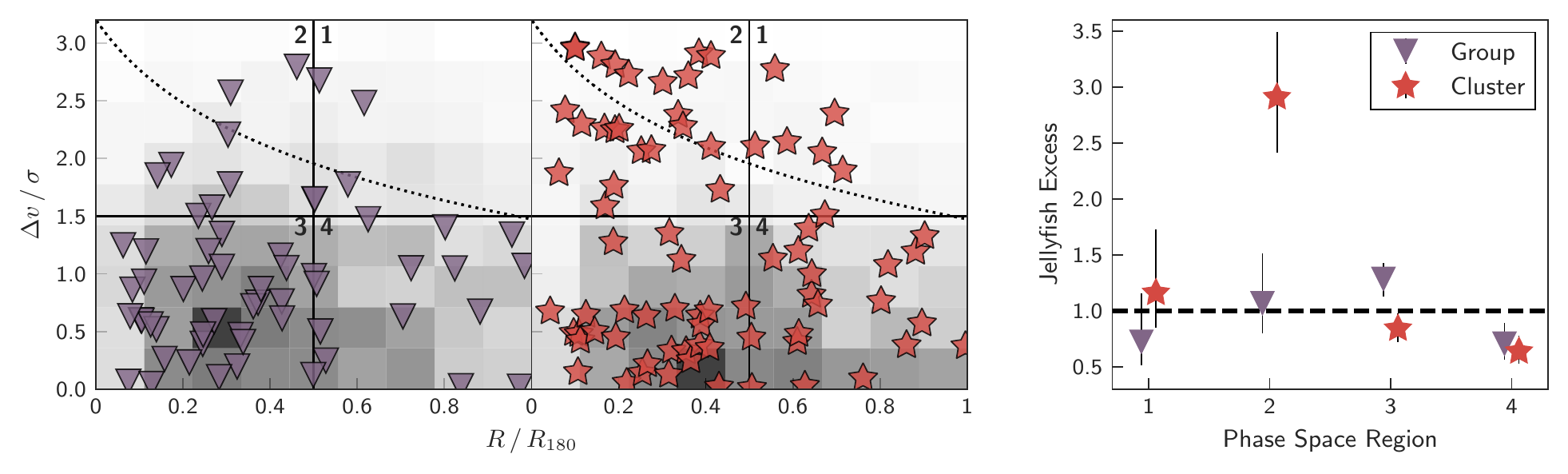}
    \caption{\textit{Left:}  Projected phase space diagrams for groups and clusters.  Data markers correspond to jellyfish galaxies in groups (purple triangles) and clusters (red stars), and the background 2D histograms show the phase space distribution for SDSS group galaxies and SDSS cluster galaxies in their respective panels. For reference, we also show the escape velocity caustic for an NFW density profile with the dotted line \citep[e.g.][]{navarro1997,jaffe2015}.}  \textit{Right:}  The excess of jellyfish galaxies, relative to SDSS group/cluster galaxies, in each of the four phase space quadrants.  Red stars correspond to jellyfish galaxies in clusters and purple triangles show jellyfish galaxies in groups.  Error bars are $1\sigma$ statistical uncertainties following \citet{cameron2011}.
    \label{fig:phase_space}
\end{figure*}

We now consider the positions of group and cluster jellyfish galaxies in PPS (velocity offset versus projected radius).  PPS distributions contain valuable information with regard to satellite galaxy infall histories, as recent infallers are typically found at large velocity offsets and/or large projected radius whereas galaxies with long times-since-infall tend to inhabit the core of PPS at small radius and small velocity offset.
\par
In Fig.~\ref{fig:phase_space} (left) we plot the PPS distributions for jellyfish galaxies in groups (purple triangles) and jellyfish galaxies in clusters (red stars).  We also show the distribution of SDSS group/cluster star-forming galaxies as the background histogram.  As in \citetalias{roberts2021_LOFARclust}, we split PPS into four quadrants divided at $\Delta v / \sigma = 1.5$ and $R/R_{180} = 0.5$.  Just by-eye there are clear differences apparent between the group and cluster PPS distributions.  In clusters there is a substantial population of jellyfish galaxies in quadrant 2, which should contain a high fraction of galaxies on their first infall.  This population is notably missing for group jellyfish galaxies, and instead most jellyfish galaxies in groups are found at small velocity offsets and small radii.
\par
We quantify these trends in the right-hand panel of Fig.~\ref{fig:phase_space} where we plot the `excess' of jellyfish galaxies (relative to SDSS star-forming galaxies) for each of the phase space quadrants.  The jellyfish excess is defined as the fraction of the group/cluster jellyfish galaxy sample in each quadrant divided by the fraction of the group/cluster SDSS star-forming sample in each quadrant.  Functionally, this is given by
\begin{equation}
    \mathrm{Jellyfish\;excess} = \left. \left(\frac{N_\mathrm{jellyfish}^{Q_i}}{N_\mathrm{jellyfish}}\right) \;\right/\; \left(\frac{N_\mathrm{SDSS}^{Q_i}}{N_\mathrm{SDSS}}\right),
\end{equation}
\noindent where $N_\mathrm{jellyfish}^{Q_i}$ is the number of LoTSS jellyfish galaxies in each quadrant, $Q_i$, and $N_\mathrm{jellyfish}$ is the total number of LoTSS jellyfish galaxies, and similarly $N_\mathrm{SDSS}^{Q_i}$ is the number of SDSS group/cluster galaxies in each quadrant, $Q_i$, and $N_\mathrm{SDSS}$ is the total number of SDSS group/cluster galaxies.
\par
As presented in \citetalias{roberts2021_LOFARclust}, there is a clear excess of cluster jellyfish galaxies in quadrant 2, consistent with cluster galaxies experiencing strong ram pressure shortly after infall.  The same is not seen in Fig.~\ref{fig:phase_space} for jellyfish galaxies in groups.  Instead, group jellyfish have a phase space distribution much more similar to the SDSS star-forming group galaxy population, with only a small fraction of galaxies at the velocity extremes in PPS.  The different PPS distributions for group and cluster jellyfish galaxies are fully consistent with the picture suggested by the tail orientations in Fig.~\ref{fig:tail_orientation}, namely that cluster jellyfish galaxies are largely being stripped on their first infall whereas group jellyfish galaxies have longer times-since-infall and many have already passed their orbital pericentre.


\section{Galaxy star formation} \label{sec:sfr}

\begin{figure}
    \centering
    \includegraphics[width=0.9\columnwidth]{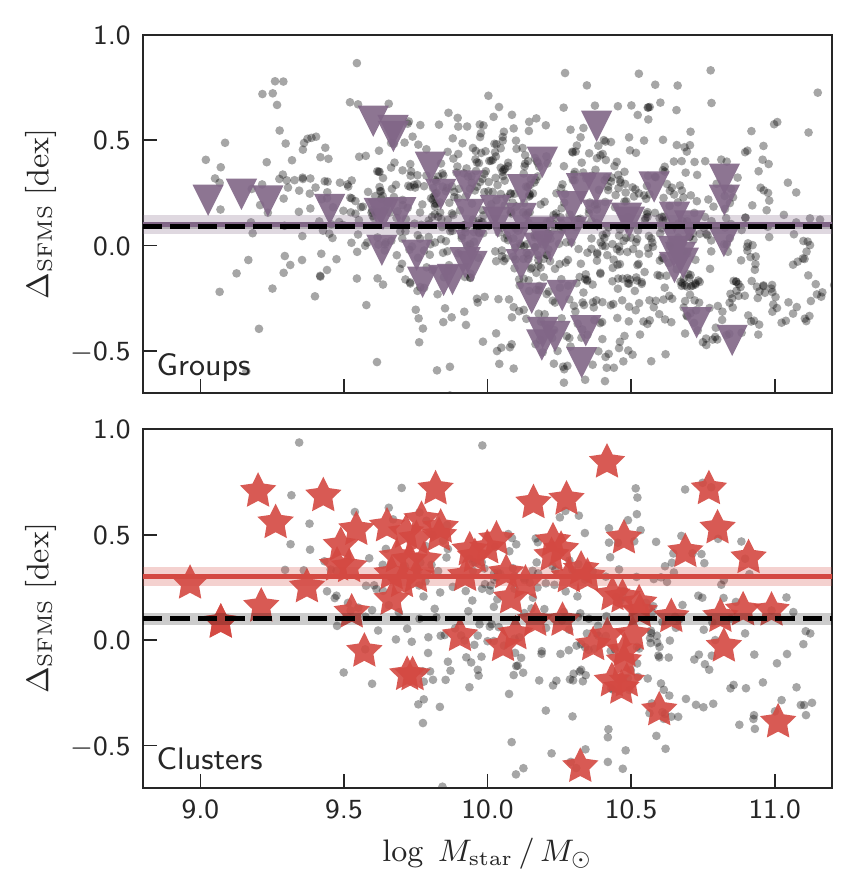}
    \caption{Offset from the star-forming main sequence (SFMS) for jellyfish galaxies in groups (top, triangles) and clusters (bottom, stars). In each panel we also show the offset from the SFMS for group/cluster LoTSS galaxies.  The SFMS relation is taken from \citetalias{roberts2021_LOFARclust} and the median offset from the SFMS is shown for jellyfish galaxies (solid line) and LoTSS galaxies (dashed line).  Shaded regions show $1\sigma$ errors on the median estimated from 5000 random bootstrap re-samplings.}
    \label{fig:sfr}
\end{figure}

Ram pressure stripping is closely tied to galaxy star formation, not only in the sense of quenching, but also through star formation enhancements (prior to substantial gas stripping) which have been predicted by simulations and observed in cluster galaxies \citep[e.g.][]{steinhauser2012,ebeling2014,vulcani2018b,ramos-martinez2018,roberts2020,troncoso-iribarren2020,durret2021}.  The origin of these star formation enhancements are often explained in terms of shocks from the ram pressure interaction which induce compression and high gas densities in the galaxy interstellar medium (ISM), in turn catalyzing strong star formation.  In groups, ram pressure is relatively weak compared to clusters, therefore it is interesting to explore whether such star formation enhancements are also present in group jellyfish galaxies.  For example, it could be that the relatively weak ram pressure in groups does not perturb the galaxy ISM as significantly as in clusters, and therefore comparable enhancements in star formation may not be expected.
\par
In Fig.~\ref{fig:sfr} we plot the offset from the SFMS for jellyfish galaxies in both groups (top, purple triangles) and clusters (bottom, red stars).  We use the best fit SFMS relation from \citetalias{roberts2021_LOFARclust}, which was derived by fitting a powerlaw relationship between SFR and stellar mass for isolated field galaxies over the same redshift range as our group/cluster samples.  As a reminder, SFRs for each galaxy are taken from the GSWLC-2 SED fitting catalogue (see Sect.~\ref{sec:galaxy_sample}, \citealt{salim2016,salim2018}).  The offset from the SFMS for each jellyfish galaxy is shown with the data markers in Fig.~\ref{fig:sfr} (Groups: purple triangles, Clusters: red stars).  We also plot offsets from the SFMS for each group/cluster LoTSS galaxy in the corresponding panel with the grey data points.  Finally, the median SFMS offset for jellyfish galaxies and for LoTSS galaxies are shown in each panel with the solid line.  Jellyfish galaxies in clusters are systematically above the SFMS but the same is not apparent for group jellyfish (at most jellyfish galaxies in groups are marginally above the SFMS). With these trends in mind, it is important to consider the selection effects given our prerequisite that galaxies be detected at 144 MHz.  144 MHz emission is a good tracer of galaxy star formation \citep{gurkan2018,smith2021}, therefore galaxies selected according to 144 MHz emission will tend to have high SFRs, which will contribute to the positive offsets from the SFMS in Fig.~\ref{fig:sfr}. This is particularly true for low-mass galaxies as can be seen in Fig.~\ref{fig:sfr} where the majority of low-mass galaxies ($M_\mathrm{star} \lesssim 10^{10}\,\mathrm{M_\odot}$) have positive SFMS offsets. This reflects the fact that in order to be detected at $144\,\mathrm{MHz}$, low-mass galaxies need to have SFRs near or above the SFMS. Conversely, given the correlation between stellar mass and SFR, high-mass galaxies can have SFRs that are below the SFMS but still high enough to be detected at $144\,\mathrm{MHz}$. This emphasizes the importance of constructing a comparison sample of `normal' LoTSS galaxies that are subject to the same selection effects as the LoTSS jellyfish galaxies. To properly gauge the enhancement (or lackthereof) of SFR in jellyfish galaxies, we show the median SFMS offset for non-jellyfish LoTSS group/cluster galaxies with the dashed lines in Fig.~\ref{fig:sfr}.  For jellyfish galaxies in groups, the offset from the SFMS is consistent with what is seen from the non-jellyfish LoTSS galaxy sample.  We do not find evidence for a true enhancement in SFR for group jellyfish galaxies, and the positive offsets from the SFMS are consistent with the selection function of the LoTSS galaxy sample.  Conversely, as shown in \citetalias{roberts2021_LOFARclust} (and reproduced in Fig.~\ref{fig:jellyfish_Mh}), cluster jellyfish galaxies have SFRs which are enhanced relative to the SFMS but also are enhanced relative to LoTSS cluster galaxies.  Therefore there is evidence for SFR enhancements in cluster jellyfish galaxies that are not present for jellyfish galaxies in groups.
\par
Previous results finding observational evidence for enhanced SFRs in RPS galaxies have focused on the galaxy cluster environment \citep{ebeling2014,poggianti2016,vulcani2018b,roberts2020,durret2021}, and these enhancements are also seen in the cluster sample from \citetalias{roberts2021_LOFARclust} and reproduced here (Fig.~\ref{fig:sfr}, bottom).  That said, there has been very little work probing the SFRs of galaxies undergoing RPS in lower mass groups.  The results of this work are qualitatively consistent with \citet{roberts2021_CFIS}, who show that ram pressure candidate galaxies in groups (identified from rest-frame optical imaging) have SFRs which are only marginally enhanced compared to much clearer SFR enhancements for the ram pressure candidates in their sample hosted by clusters.  The origin of this difference between groups and clusters is not immediately clear, although as previously mentioned, it is possible that the more intense ram pressure in clusters can more strongly perturb the ISM in galaxies, leading to enhanced gas densities and increased star formation.  This is largely speculative at this point, though this could be tested with observations of cold gas in both group and cluster jellyfish galaxies, and also through comparisons to hydrodynamic simulations of group and cluster galaxies.


\section{Discussion} \label{sec:disc_conc}

Here we have presented a contrast between the properties of LoTSS 144 MHz jellyfish galaxies in groups compared to clusters.  We find clear differences between the two environments, all of which are consistent with a picture where galaxies in groups are less strongly affected by RPS than galaxies in clusters.  Given the higher ICM densities and velocity dispersions in clusters, less efficient RPS stripping in groups is a natural expectation, however this is one of the first studies to show such clear evidence for this picture. Below, we discuss our conclusions in the context of a simple toy model for RPS, as well as the implications of these results for the pre-processing of galaxies prior to cluster infall.

\subsection{Ram pressure toy model} \label{sec:disc_toymodel}

A primary interpretation of the results from this work is that RPS is a more rapid process in clusters than groups.  This can be seen from the tail orientations in Fig.~\ref{fig:tail_orientation} or the phase space diagrams in Fig.~\ref{fig:phase_space}, both of which are consistent with cluster jellyfish galaxies being primarily on first infall (before pericentre) whereas many group jellyfish galaxies are consistent with backsplashing orbits after a pericentric passage.  The crux of this interpretation relies on the strength of RPS being relatively modest in  groups, such that galaxies are not completely stripped on their first infall.  In this section we present a very simple toy model of RPS in order to show that qualitative expectations from such a model are consistent with this picture.  We note that this simple approach is not a complete description of RPS, instead, it is meant to show the qualitative variations in RPS timescales between low-mass groups and massive clusters.
\par
We follow many previous works and model ram pressure stripping through the balance between the strength of ram pressure and the gravitational potential of a galaxy \citep[e.g.][]{gunn1972,rasmussen2008,jaffe2018,roberts2019}.  We take an extremely simple galaxy model consisting of a thin exponential stellar disk and a thin exponential gas disk, each with different scale lengths. We note that an exponential disk distribution should also be, roughly, true of galaxy \textsc{Hii} regions that are likely the source of stripped plasma observed in the jellyfish galaxies in this work.  The ram pressure, $P_\mathrm{ram}$, and the galaxy anchoring force, $\Pi$, are then given by
\begin{align}
    &P_\mathrm{ram}(R) = \rho_\mathrm{ICM}(R)\,v^2 \\
    &\Pi(r) = 2 \pi G \Sigma_\star(r) \Sigma_\mathrm{gas}(r)
\end{align}
\noindent
For a given value of $\rho_\mathrm{ICM}$ and $v$, one can define a `stripping radius', $r_\mathrm{strip}$, within a model galaxy corresponding to the largest galactocentric radius where the inequality,
\begin{equation*}
\rho_\mathrm{ICM}(R)\,v^2 > 2 \pi G \Sigma_\star(r) \Sigma_\mathrm{gas}(r)
\end{equation*}
\noindent
is satisfied.  For our model we assume that for a given $\rho_\mathrm{ICM}$ and $v$, the \textsc{Hi} gas disk is truncated at $r=r_\mathrm{strip}$ due to ram pressure, such that any gas located beyond $r_\mathrm{strip}$ is completely removed from the galaxy.
\par
\begin{figure}
    \centering
    \includegraphics[width=\columnwidth]{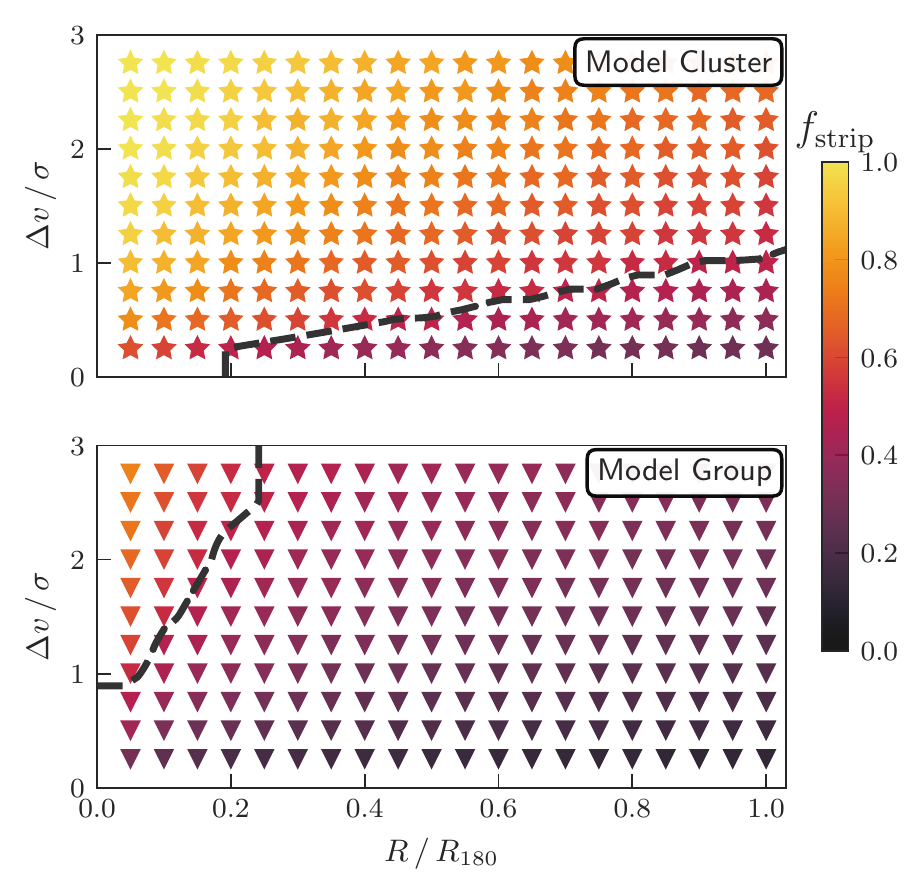}
    \caption{Results from the ram pressure stripping toy model for our model cluster (top) and model group (bottom).  The grid points are coloured by the fraction of gas mass stripped by the ram pressure toy model ($f_\mathrm{strip}$), and the dashed contour corresponds to a value of $f_\mathrm{strip}=0.5$.  All values of $f_\mathrm{strip}$ are calculated for the model galaxy with the parameters described in Table~\ref{tab:toy_model}.}
    \label{fig:toy_infall}
\end{figure}

For this toy model we consider a model galaxy orbiting through a model galaxy cluster and a model galaxy group, and the relevant parameter values are listed in Table~\ref{tab:toy_model}.  We model our toy cluster after the Coma Cluster and we model our toy group after the NGC 4636 group.  We select these two examples to use because they are among the most massive (Coma, $M_{180} \sim 2\times10^{15}\,\mathrm{M_\odot}$) and least massive (NGC 4636 Grp, $M_{180} \sim 2\times10^{13}\,\mathrm{M_\odot}$) systems in the \citet{chen2007} sample, and roughly span the entire halo mass range from this work.  Therefore the differences in Fig.~\ref{fig:toy_infall} can be thought of as the broad differences expected between the least massive and most massive systems in our sample.
\begin{table}
    \centering
    \caption{Ram Pressure Model Parameters}
    \begin{threeparttable}
    \begin{tabular}{r c c c}
        \toprule
        & Model Cluster & Model Group & Ref. \\
        \midrule
        Modelled After: & Coma & NGC 4636 & \\
        $\rho_0 \; \mathrm{(g\,cm^{-3})}$: & $5.0 \times 10^{-27}$ & $2.8 \times 10^{-26}$ & a \\
        $R_c \; \mathrm{(kpc)}$: & 343 & 6 & a \\
        $\beta$: & 0.654 & 0.491 & a \\
        $R_{180} \; \mathrm{(kpc)}$: & 2982 & 803 & b,c \\
        $\sigma_v \; \mathrm{(km\,s^{-1})}$: & 1082 & 284 & c,d \\
        \midrule
        & Model Galaxy && \\
        \midrule
        $M_\mathrm{star} \; \mathrm{(M_\odot)}$: & $1 \times 10^{10}$ && \\
        $R_{d,\star} \; \mathrm{(kpc)}$: & 2.0 && e.g. e,f  \\
        $M_\mathrm{gas} \; \mathrm{(M_\odot)}$: & $3.3 \times 10^{9}$ && g \\
        $R_{d,gas} \; \mathrm{(kpc)}$: & 3.4 && h \\
        \bottomrule
    \end{tabular}
    \begin{tablenotes}
    \small
    \item[a] \citet{chen2007}
    \item[b] \citet{kubo2007}
    \item[c] \citet{osmond2004}
    \item[d] \citet{colless1996}
    \item[e] \citet{fathi2010}
    \item[f] \citet{demers2019}
    \item[g] \citet{brown2015}
    \item[h] \citet{cayatte1994}
    \end{tablenotes}
    \end{threeparttable}
    \label{tab:toy_model}
\end{table}

\par
In Fig.~\ref{fig:toy_infall} we show the fraction of stripped \textsc{Hi} mass, $f_\mathrm{strip}$ (colourbar), as a function of position in projected phase space, for the model cluster (top) and model group (bottom).  The dashed contour in each panel corresponds to $f_\mathrm{strip} = 0.5$.  The RPS predictions clearly differ between the model group and cluster, which is driven both by the different ICM density profiles and the different velocity dispersions between the two systems.  According to this simple model, substantial fractions (>50\%) of galaxy gas reserves are stripped shortly after passing $R_{180}$ in clusters.  For groups this is not the case, and the only region of phase space for groups where $f_\mathrm{strip} > 0.5$ is at very small radii and very large velocity offsets.  This shows how RPS can be less efficient in groups relative to clusters.  The differences between groups and clusters from this toy model are consistent with our interpretation of the observed trends in this work; namely, that galaxies in clusters are primarily stripped on their first infall whereas galaxies in groups can maintain significant gas reserves beyond first pericentre.
\par
We reiterate that this is a very simplistic treatment of ram pressure stripping, and is not meant to realistically capture the details of stripping in groups and clusters.  While we take a single group model for illustrative purposes, in reality there is likely significant scatter in the ICM densities for different groups.  This scatter in ICM density, and in particular whether a group lies on the high or low density end, likely also plays an important role in determining the efficiency of ram pressure stripping in such low-mass environments.  Additionally, this model likely overestimates the amount of gas stripping somewhat, given that we do not include any contributions from a stellar bulge or dark matter halo to the galaxy restoring potential and that we do not include any contribution from the more densely bound molecular component to the total gas mass.  All said, the purpose of this exercise is to illustrate the broad differences in the efficiency of RPS between the group and cluster environment.  Furthermore, to show that galaxies in clusters being stripped shortly after infall is a reasonable expectation, as is stripping timescales in groups extending beyond the first passage of pericentre.

\subsection{Implications for pre-processing} \label{sec:disc_preproc}

The presence of jellyfish galaxies in groups also has important implications for pre-processing, as RPS is likely relevant for the quenching of satellite star formation in the group regime (albeit less efficiently than for clusters).  There have been a number of estimates in literature for the fraction of cluster galaxies that have been pre-processed, in other words, the fraction of galaxies on the cluster red sequence that were quenched in a lower mass group and subsequently accreted onto the cluster as a passive galaxy.  As many as half of present day cluster galaxies may have been accreted as a group member \citep[e.g.][]{mcgee2009,delucia2012,bahe2013,hou2014}, though not all of those galaxies will have been pre-processed in the sense that not all galaxies infalling as group members will be quenched.  More direct constraints on the fraction of pre-processed galaxies have been made, and typically fall between $\sim$10\% and $\sim$30\% \citep{haines2015,roberts2017,olave2018,vanderburg2018,roberts2019}.  Depending on the group mass, we find that between 5\% and 15\% of LoTSS-detected star-forming galaxies show signs of RPS in the radio continuum (Fig.~\ref{fig:jellyfish_Mh}).  Due to the LoTSS sensitivity limits we are not sensitive to the low SFRs typical of low-mass galaxies around $\sim\!10^9\,\mathrm{M_\odot}$ (assuming a typical SFMS). These low-mass galaxies are expected to be strongly impacted by RPS \citep[e.g.][]{fillingham2015,roberts2019,yun2019,baxter2021,roberts2021_CFIS}, therefore the fractions in Fig.~\ref{fig:jellyfish_Mh} would likely be larger if we could probe down to lower stellar masses.  The fractions in Fig.~\ref{fig:jellyfish_Mh} also only account for galaxies which currently show morphological features consistent with RPS, and do not include `post-stripping' galaxies with symmetric gas disks that have already been truncated by ram pressure \citep[e.g.][]{sengupta2007,vollmer2007,jaffe2018,vulcani2018}.
\par
If RPS stripping is contributing to pre-processing in groups, this implies that galaxies infalling onto clusters as part of a group should already be \textsc{Hi} deficient, to some extent, relative to field galaxies.  This is consistent with \textsc{Hi} observations from the BUDHIES survey which find \textsc{Hi} deficient galaxies in group-mass substructures surrounding the Abell 963 cluster \citep{jaffe2016}, as well as MeerKAT observations finding \textsc{Hi} deficient galaxies in the Fornax A group \citep{kleiner2021}.  Other works have also reported evidence for \textsc{Hi} deficient galaxies in groups \citep[e.g.][]{huchtmeier1997,verdes-montenegro2001,denes2016,brown2017}, which based on the results of this work could be driven by RPS.
\par
Beyond gravitationally bound groups, galaxies may also be pre-processed in cosmic filaments prior to cluster infall.  This can been seen by the fact that the fraction of red, quenched galaxies increases toward the central spine of filaments \citep[e.g.][]{kuutma2017,malavasi2017,kraljic2018,salerno2019}. Recently, \citet{bonjean2018} have shown that galaxies in the filament bridge between the Abell 399 and Abell 401 clusters have indistinguishable properties (i.e.\ early-type, passively evolving) from galaxies within the clusters. This suggests that galaxy properties are impacted by these dense filamentary environments. It has been suggested that ram pressure  could affect galaxies even within cosmic filaments \citep{benitez-llambay2013,vulcani2018}, though given the relatively low gas densities in filaments compared to groups and clusters \citep[e.g.][]{edwards2010,eckert2015,tanimura2020}, the efficiency of RPS in such environments is likely low.  It is also likely that the gas and galaxies in filaments are moving more coherently than in groups or clusters, meaning that the relative velocities could be lower and less conducive to RPS.  While this it is not the focus of this work, it may be possible to constrain the presence, or lack thereof, of RPS in cosmic filaments with LoTSS.  LoTSS DR2 covers $\sim\!5700\,\mathrm{deg^2}$ in the northern sky at both high and low galactic latitude. With such a wide area of the extragalactic sky it is possible to probe the properties of filament galaxies in a statistical fashion.  Given a sample of galaxies in filaments, for example identified from SDSS spectroscopy or from filament bridges between nearby clusters, a search for potential jellyfish galaxies could then be done with similar methods as this work.

\section{Summary} \label{sec:summary}

In this work we present a search for radio continuum jellyfish galaxies with LOFAR across a sample of $\sim$500 low redshift galaxy groups ($10^{12.5} < M_\mathrm{group} < 10^{14}\,\mathrm{M_\odot}$).  We also incorporate the sample of radio continuum jellyfish galaxies in clusters from \citetalias{roberts2021_LOFARclust}, allowing us to contrast the properties of jellyfish galaxies in groups and clusters across three decades in halo mass.  The main conclusions from this work are summarized below.

\begin{enumerate}
    \itemsep0.5em
    \item The frequency of jellyfish galaxies is highest in clusters and lowest in low-mass groups (Fig.~\ref{fig:jellyfish_Mh}).

    \item We find evidence for weaker ram pressure stripping in groups relative to clusters.  Many jellyfish galaxies in groups are consistent with having already passed pericentre, which does not seem to be the case for jellyfish galaxies in clusters (Figs \ref{fig:tail_orientation} \& \ref{fig:phase_space}).

    \item Unlike jellyfish galaxies in clusters, jellyfish galaxies in groups do not have systematically enhanced star formation rates (Fig.~\ref{fig:sfr}).
\end{enumerate}
\noindent
The results of this work highlight that ram pressure stripping of galaxies is occurring in groups, and that there are interesting differences between the properties of jellyfish galaxies in groups and clusters.  Moving forward it will be important to obtain detailed, multiwavelength observations of group jellyfish galaxies (e.g. optical IFU, \textsc{Hi}, molecular gas) as has already been done for such galaxies in clusters \citep[e.g.][]{chung2007,poggianti2017,jachym2019,moretti2020}.  This will aid in understanding the similarities and differences between the impact of ram pressure on galaxy evolution in both the group and cluster regimes.

\begin{acknowledgements}
IDR and RJvW acknowledge support from the ERC Starting Grant Cluster Web 804208. SLM acknowledges support from STFC through grant number ST/N021702/1. AB acknowledges support from the VIDI research programme with project number 639.042.729, which is financed by the Netherlands Organisation for Scientific Research (NWO). AI acknowledges the Italian PRIN-Miur 2017 (PI A. Cimatti).
\par
This paper is based on data obtained with the International LOFAR Telescope (ILT). LOFAR \citep{vanhaarlem2013} is the LOw Frequency ARray designed and constructed by ASTRON. It has observing, data processing, and data storage facilities in several countries, which are owned by various parties (each with their own funding sources) and are collectively operated by the ILT foundation under a joint scientific policy. The ILT resources have benefited from the following recent major funding sources: CNRS-INSU, Observatoire de Paris and Universit\'e d'Orl\'eans, France; BMBF, MIWF-NRW, MPG, Germany; Science Foundation Ireland (SFI), Department of Business, Enterprise and Innovation (DBEI), Ireland; NWO, The Netherlands; The Science and Technology Facilities Council, UK; Ministry of Science and Higher Education, Poland; The Istituto Nazionale di Astrofisica (INAF), Italy. This research made use of the Dutch national e-infrastructure with support of the SURF Cooperative (e-infra 180169) and the LOFAR e-infra group. The J\"ulich LOFAR Long Term Archive and the GermanLOFAR network are both coordinated and operated by the J\"ulich Supercomputing Centre (JSC), and computing resources on the supercomputer JUWELS at JSC were provided by the Gauss Centre for Supercomputinge.V. (grant CHTB00) through the John von Neumann Institute for Computing (NIC). This research made use of the University of Hertfordshire high-performance computing facility (\url{http://uhhpc. herts.ac.uk}) and the LOFAR-UK computing facility located at the University of Hertfordshire and supported by STFC [ST/P000096/1], and of the Italian LOFAR IT computing infrastructure supported and operated by INAF, and by the Physics Department of Turin University (under an agreement with Consorzio Interuniversitario per la Fisica Spaziale) at the C3S Supercomputing Centre, Italy.
\par
The Pan-STARRS1 Surveys (PS1) and the PS1 public science archive have been made possible through contributions by the Institute for Astronomy, the University of Hawaii, the Pan-STARRS Project Office, the Max-Planck Society and its participating institutes, the Max Planck Institute for Astronomy, Heidelberg and the Max Planck Institute for Extraterrestrial Physics, Garching, The Johns Hopkins University, Durham University, the University of Edinburgh, the Queen's University Belfast, the Harvard-Smithsonian Center for Astrophysics, the Las Cumbres Observatory Global Telescope Network Incorporated, the National Central University of Taiwan, the Space Telescope Science Institute, the National Aeronautics and Space Administration under Grant No. NNX08AR22G issued through the Planetary Science Division of the NASA Science Mission Directorate, the National Science Foundation Grant No. AST-1238877, the University of Maryland, Eotvos Lorand University (ELTE), the Los Alamos National Laboratory, and the Gordon and Betty Moore Foundation.
\end{acknowledgements}

%
%

\bibliographystyle{aa}
\bibliography{main}

\begin{appendix}
\onecolumn
\section{Jellyfish galaxies} \label{sec:img_appendix}

\input{RPtable}

\begin{figure*}
    \centering
    \includegraphics[width=\textwidth]{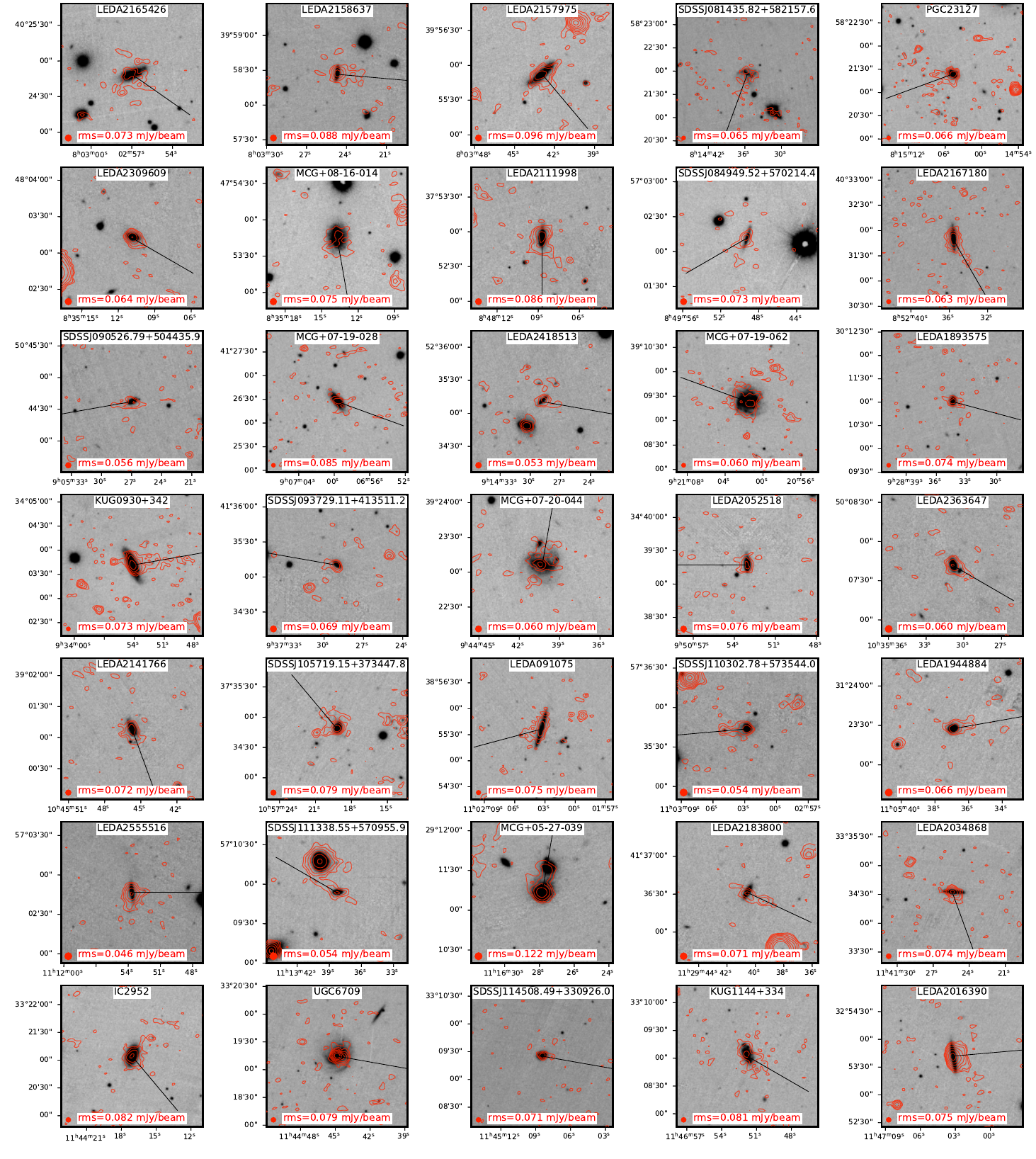}
    \caption{$100\,\mathrm{kpc} \times 100\,\mathrm{kpc}$ Pan-STARRs $g$-band, LOFAR 144 MHz overlay images for LoTSS jellyfish galaxies.  144 MHz contours are logarithmically spaced starting at $2 \times$ the rms, with the rms level labeled in each cutout panel, and the $6''$ LoTSS beam is shown in the lower left of each panel.  The black line in each panel denotes the tail direction estimated in Sect.~\ref{sec:orbital_hist}.}
    \label{fig:panel_img1}
\end{figure*}

\begin{figure*}
    \centering
    \includegraphics[width=\textwidth]{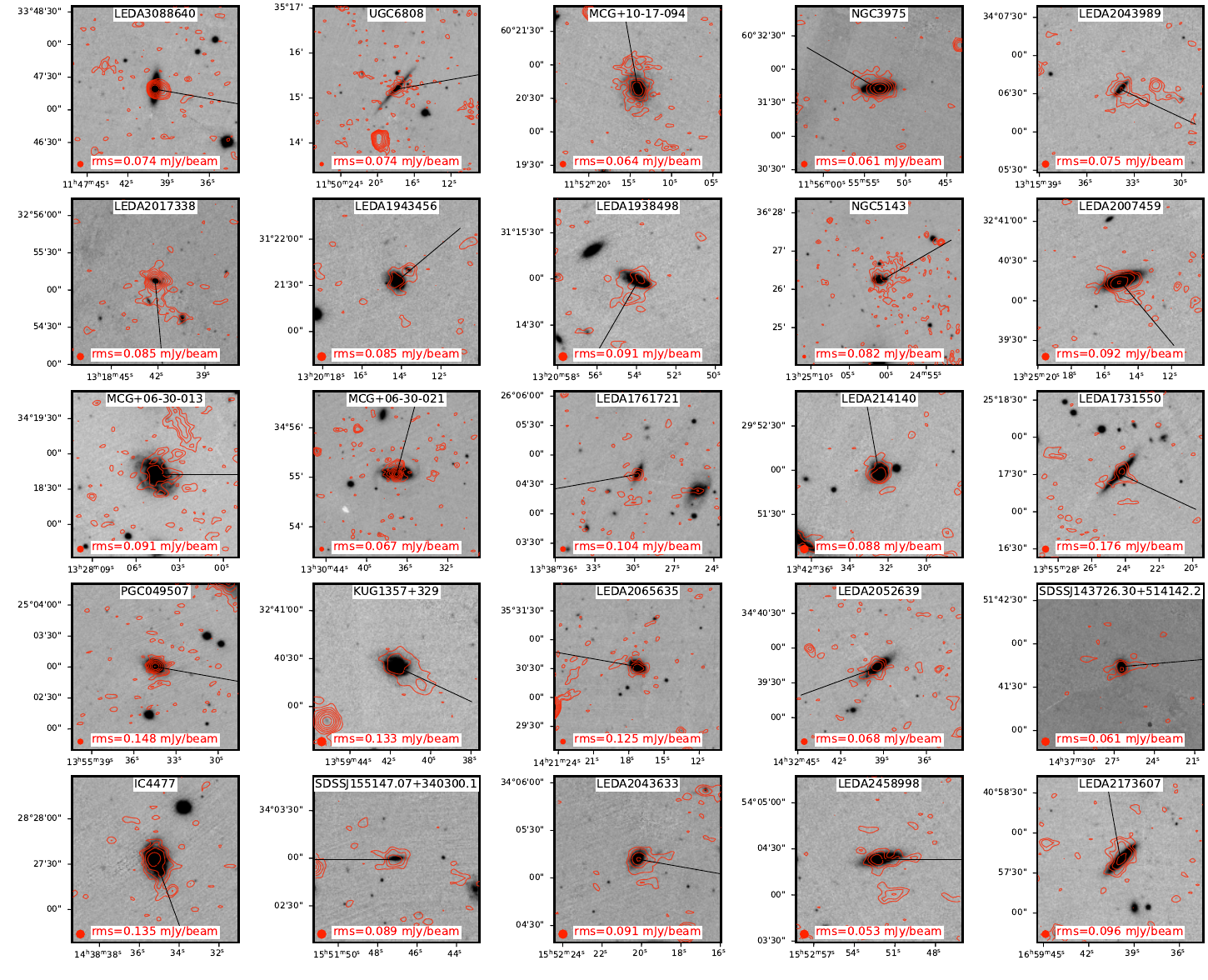}
    \caption{$100\,\mathrm{kpc} \times 100\,\mathrm{kpc}$ Pan-STARRs $g$-band, LOFAR 144 MHz overlay images for LoTSS jellyfish galaxies.  144 MHz contours are logarithmically spaced starting at $2 \times$ the rms, with the rms level labeled in each cutout panel, and the $6''$ LoTSS beam is shown in the lower left of each panel.  The black line in each panel denotes the tail direction estimated in Sect.~\ref{sec:orbital_hist}.}
    \label{fig:panel_img2}
\end{figure*}

\end{appendix}

\end{document}

%% file: RPtable.tex
\begin{ThreePartTable}
\begin{TableNotes}
\footnotesize
\item [a] \citet{salim2016,salim2018}
\item [b] 144 MHz flux density from LoTSS DR2 source catalog
\end{TableNotes}
\begin{longtable}{l c c c c c c c c}
\caption{Group jellyfish galaxies} \\
\toprule
Name & RA & Dec & $z$ & $\log M_\mathrm{star}$\tnote{a} & $\log \mathrm{SFR}$\tnote{a} & $R/R_{180}$ & $\Delta v / \sigma$ & $f_\mathrm{144}$\tnote{b} \\
& [deg] & [deg] & & [$\mathrm{M_\odot}$] & [$\mathrm{M_\odot\,yr^{-1}}$] & & & [mJy] \\
\midrule
\endfirsthead
\toprule
Name & RA & Dec & $z$ & $\log M_\mathrm{star}$\tnote{a} & $\log \mathrm{SFR}$\tnote{a} & $R/R_{180}$ & $\Delta v / \sigma$ & $f_\mathrm{144}$\tnote{b} \\
& [deg] & [deg] & & [$\mathrm{M_\odot}$] & [$\mathrm{M_\odot\,yr^{-1}}$] & & & [mJy] \\
\midrule
\endhead
\midrule
\multicolumn{9}{r}{{Continued on next page}} \\
\bottomrule
\endfoot
\bottomrule
\insertTableNotes
\endlastfoot
LEDA2165426 & 120.7372 & 40.4134 & 0.0426 & 10.4 & 0.18 & 0.07 & 0.07 & 6.0 $\pm$ 0.9 \\
LEDA2158637 & 120.8532 & 39.9742 & 0.0419 & 9.7 & 0.16 & 0.27 & 0.59 & 2.6 $\pm$ 0.3 \\
LEDA2157975 & 120.9291 & 39.9314 & 0.0423 & 10.3 & 0.17 & 0.29 & 1.06 & 7.1 $\pm$ 0.6 \\
SDSSJ081435.82+582157.6 & 123.6493 & 58.3661 & 0.0293 & 9.0 & -0.56 & 0.27 & 1.57 & 2.9 $\pm$ 0.7 \\
PGC23127 & 123.7712 & 58.3571 & 0.0260 & 9.1 & -0.46 & 0.24 & 0.96 & 4.0 $\pm$ 0.4 \\
LEDA2309609 & 128.7962 & 48.0541 & 0.0413 & 10.4 & 0.31 & 0.30 & 2.21 & 2.3 $\pm$ 0.2 \\
MCG+08-16-014 & 128.8070 & 47.8957 & 0.0429 & 10.7 & 0.14 & 0.52 & 0.50 & 0.8 $\pm$ 0.2 \\
LEDA2111998 & 132.0368 & 37.8823 & 0.0397 & 10.0 & -0.01 & 0.46 & 2.79 & 6.1 $\pm$ 0.5 \\
SDSSJ084949.52+570214.4 & 132.4564 & 57.0374 & 0.0417 & 9.6 & -0.43 & 0.28 & 0.10 & 1.3 $\pm$ 0.2 \\
LEDA2167180 & 133.1490 & 40.5311 & 0.0286 & 10.1 & -0.21 & 0.58 & 1.78 & 4.9 $\pm$ 0.4 \\
SDSSJ090526.79+504435.9 & 136.3617 & 50.7433 & 0.0378 & 9.4 & -0.35 & 0.50 & 0.12 & 2.3 $\pm$ 0.4 \\
MCG+07-19-028 & 136.7483 & 41.4406 & 0.0261 & 9.8 & -0.50 & 0.06 & 1.25 & 1.9 $\pm$ 0.4 \\
LEDA2418513 & 138.6206 & 52.5866 & 0.0398 & 9.9 & -0.42 & 0.11 & 0.59 & 1.0 $\pm$ 0.2 \\
MCG+07-19-062 & 140.2570 & 39.1566 & 0.0298 & 10.5 & 0.22 & 0.62 & 2.48 & 9.7 $\pm$ 0.8 \\
LEDA1893575 & 142.1436 & 30.1838 & 0.0290 & 9.9 & -0.22 & 0.29 & 1.35 & 1.9 $\pm$ 0.3 \\
KUG0930+342 & 143.4763 & 34.0619 & 0.0265 & 10.3 & -0.27 & 0.17 & 1.94 & 5.0 $\pm$ 0.3 \\
SDSSJ093729.11+413511.2 & 144.3713 & 41.5865 & 0.0412 & 9.8 & -0.04 & 0.38 & 0.87 & 1.1 $\pm$ 0.2 \\
MCG+07-20-044 & 146.1681 & 39.3853 & 0.0406 & 10.8 & 0.32 & 0.51 & 0.91 & 0.9 $\pm$ 0.2 \\
LEDA2052518 & 147.7212 & 34.6549 & 0.0399 & 9.7 & -0.21 & 0.53 & 0.25 & 1.6 $\pm$ 0.3 \\
LEDA2363647 & 158.8797 & 50.1290 & 0.0463 & 9.8 & 0.06 & 0.50 & 0.98 & 2.0 $\pm$ 0.3 \\
LEDA2141766 & 161.4409 & 39.0194 & 0.0359 & 9.6 & 0.16 & 0.43 & 0.76 & 3.0 $\pm$ 0.4 \\
SDSSJ105719.15+373447.8 & 164.3298 & 37.5800 & 0.0359 & 9.7 & 0.13 & 0.13 & 0.05 & 3.1 $\pm$ 0.3 \\
LEDA091075 & 165.5138 & 38.9270 & 0.0303 & 10.2 & -0.49 & 0.32 & 0.20 & 5.3 $\pm$ 0.6 \\
SDSSJ110302.78+573544.0 & 165.7616 & 57.5956 & 0.0475 & 10.2 & -0.48 & 0.84 & 0.03 & 0.9 $\pm$ 0.1 \\
LEDA1944884 & 166.4037 & 31.3906 & 0.0457 & 9.9 & -0.32 & 0.31 & 2.57 & 1.5 $\pm$ 0.2 \\
LEDA2555516 & 167.9746 & 57.0468 & 0.0479 & 10.2 & -0.35 & 0.13 & 0.56 & 2.0 $\pm$ 0.2 \\
SDSSJ111338.55+570955.9 & 168.4107 & 57.1655 & 0.0445 & 9.6 & -0.26 & 0.51 & 2.68 & 1.0 $\pm$ 0.2 \\
MCG+05-27-039 & 169.1161 & 29.1871 & 0.0489 & 10.6 & 0.44 & 0.20 & 0.87 & 5.4 $\pm$ 0.5 \\
LEDA2183800 & 172.4184 & 41.6083 & 0.0442 & 10.3 & -0.39 & 0.42 & 1.16 & 0.8 $\pm$ 0.2 \\
LEDA2034868 & 175.3559 & 33.5756 & 0.0331 & 10.2 & -0.55 & 0.26 & 1.20 & 3.7 $\pm$ 0.3 \\
IC2952 & 176.0714 & 33.3514 & 0.0322 & 10.3 & 0.05 & 0.25 & 0.39 & 5.4 $\pm$ 0.5 \\
UGC6709 & 176.1870 & 33.3210 & 0.0311 & 10.6 & 0.17 & 0.12 & 1.19 & 10.7 $\pm$ 0.2 \\
SDSSJ114508.49+330926.0 & 176.2854 & 33.1573 & 0.0331 & 9.9 & 0.05 & 0.16 & 0.26 & 1.0 $\pm$ 0.2 \\
KUG1144+334 & 176.7161 & 33.1510 & 0.0350 & 9.6 & -0.26 & 0.50 & 1.65 & 9.9 $\pm$ 1.1 \\
LEDA2016390 & 176.7639 & 32.8953 & 0.0313 & 9.9 & -0.32 & 0.72 & 1.05 & 23.1 $\pm$ 0.6 \\
LEDA3088640 & 176.9170 & 33.7890 & 0.0309 & 10.7 & 0.22 & 0.96 & 1.34 & 23.8 $\pm$ 0.3 \\
UGC6808 & 177.5756 & 35.2541 & 0.0213 & 10.3 & -0.55 & 0.50 & 1.65 & 5.1 $\pm$ 0.6 \\
MCG+10-17-094 & 178.0600 & 60.3447 & 0.0342 & 10.7 & 0.10 & 0.88 & 0.67 & 4.9 $\pm$ 0.3 \\
NGC3975 & 178.9737 & 60.5294 & 0.0329 & 10.5 & 0.23 & 0.08 & 0.63 & 13.6 $\pm$ 1.0 \\
LEDA2043989 & 198.8911 & 34.1093 & 0.0381 & 9.8 & -0.38 & 0.98 & 1.07 & 5.0 $\pm$ 0.5 \\
LEDA2017338 & 199.6759 & 32.9187 & 0.0359 & 10.1 & -0.10 & 0.80 & 1.40 & 10.5 $\pm$ 0.3 \\
LEDA1943456 & 200.0593 & 31.3589 & 0.0465 & 10.2 & -0.02 & 0.33 & 0.47 & 1.7 $\pm$ 0.4 \\
LEDA1938498 & 200.2247 & 31.2495 & 0.0452 & 10.1 & 0.01 & 0.62 & 1.47 & 3.3 $\pm$ 0.5 \\
NGC5143 & 201.2553 & 36.4372 & 0.0195 & 9.2 & -0.44 & 0.44 & 1.05 & 16.6 $\pm$ 1.2 \\
LEDA2007459 & 201.3132 & 32.6711 & 0.0398 & 10.8 & 0.52 & 0.34 & 0.41 & 7.8 $\pm$ 0.5 \\
MCG+06-30-013 & 202.0191 & 34.3115 & 0.0360 & 10.7 & 0.34 & 0.08 & 0.86 & 9.0 $\pm$ 1.2 \\
MCG+06-30-021 & 202.6540 & 34.9174 & 0.0256 & 10.7 & 0.33 & 0.24 & 1.50 & 6.1 $\pm$ 0.3 \\
LEDA1761721 & 204.6236 & 26.0776 & 0.0294 & 9.8 & -0.45 & 0.97 & 0.02 & 2.1 $\pm$ 0.4 \\
LEDA214140 & 205.6353 & 29.8660 & 0.0471 & 10.3 & 0.28 & 0.14 & 1.85 & 3.2 $\pm$ 0.4 \\
LEDA1731550 & 208.8510 & 25.2914 & 0.0370 & 10.9 & -0.13 & 0.37 & 0.75 & 7.0 $\pm$ 1.1 \\
PGC049507 & 208.8933 & 25.0498 & 0.0290 & 10.4 & 0.60 & 0.31 & 1.78 & 20.0 $\pm$ 1.2 \\
KUG1357+329 & 209.9239 & 32.6738 & 0.0498 & 10.2 & 0.32 & 0.70 & 0.63 & 7.6 $\pm$ 1.2 \\
LEDA2065635 & 215.3225 & 35.5094 & 0.0298 & 10.2 & -0.06 & 0.82 & 1.05 & 4.7 $\pm$ 0.7 \\
LEDA2052639 & 218.1638 & 34.6623 & 0.0344 & 10.0 & -0.06 & 0.36 & 0.72 & 2.4 $\pm$ 0.3 \\
SDSSJ143726.30+514142.2 & 219.3596 & 51.6951 & 0.0443 & 9.9 & -0.08 & 0.43 & 0.61 & 1.0 $\pm$ 0.2 \\
IC4477 & 219.6468 & 28.4594 & 0.0468 & 10.8 & 0.62 & 0.10 & 0.60 & 10.5 $\pm$ 0.8 \\
SDSSJ155147.07+340300.1 & 237.9461 & 34.0501 & 0.0496 & 10.2 & -0.10 & 0.11 & 0.93 & 2.1 $\pm$ 0.3 \\
LEDA2043633 & 238.0843 & 34.0868 & 0.0484 & 10.1 & 0.15 & 0.21 & 0.23 & 2.6 $\pm$ 0.3 \\
LEDA2458998 & 238.2182 & 54.0737 & 0.0466 & 10.7 & -0.12 & 0.14 & 0.52 & 4.5 $\pm$ 0.7 \\
LEDA2173607 & 254.9165 & 40.9614 & 0.0403 & 10.7 & 0.16 & 0.25 & 0.47 & 8.1 $\pm$ 0.6 \\
\label{tab:jellyfish_sample}
\end{longtable}
\end{ThreePartTable}